# Evidence of symmetry lowering in antiferromagnetic metal TmB$_{12}$ with dynamic charge stripes

A. Azarevich[1,2], V. Glushkov[1], S. Demishev[1], A. Bogach[1], V. Voronov[1], S. Gavrilkin[3], N. Shitsevalova[4], V. Filipov[4], S. Gabáni[5], J. Kačmarčík[5], K. Flachbart[5], N. Sluchanko[1]

[1]Prokhorov General Physics Institute of Russian Academy of Sciences, 38 Vavilov str., Moscow 119991, Russia

[2]Moscow Institute of Physics and Technology, 9 Institutskiy per., Dolgoprudny, Moscow Region, 141701, Russia

[3]Lebedev Physical Institute of RAS, 53 Leninsky Avenue, 119991 Moscow, Russia

[4]Frantsevich Institute for Problems of Materials Science, National Academy of Sciences of Ukraine, 3 Krzhizhanovsky str., Kyiv 03680, Ukraine

[5]Institute of Experimental Physics, Slovak Academy of Sciences, 47 Watsonova, 04001 Košice, Slovakia

**Abstract.**

Precise angle-resolved magnetoresistance (ARMR) and magnetization measurements have revealed (*i*) strong charge transport and magnetic anisotropy and (*ii*) emergence of a huge number of magnetic phases in the ground state of TmB$_{12}$ antiferromagnetic metal with *fcc* crystal structure and dynamic charge stripes. By analyzing the angular *H-φ* magnetic phase diagrams reconstructed from experimental angle-resolved magneto-resistance (ARMR) and magnetization data we argue that the symmetry lowering is a consequence of suppression of the indirect Ruderman–Kittel–Kasuya–Yosida (RKKY) exchange along <110> directions between nearest neighboring magnetic moments of Tm$^{3+}$ ions and subsequent redistribution of conduction electrons to quantum fluctuations of the electron density (stripes). Magnetoresistance components are discussed in terms of charge scattering on the spin density wave, itinerant ferromagnetic nano-domains and on-site Tm$^{3+}$ spin fluctuations.

## 1. Introduction.

Rare earth dodecaborides RB$_{12}$ demonstrate a vast diversity of magnetic and transport properties including ferro- and antiferromagnetic ground states, metal-insulator transition and intermediate valence (see, for review [1-3]) and superconductivity [4,5]. These compounds have a relatively simple and highly symmetrical UB$_{12}$-type *fcc* crystal structure [6], but various simultaneously active interactions in these strongly correlated electron systems make their quantitative theoretical description to be a difficult task. Indeed, first principles calculations have shown that the $B_{12}$ molecule stabilizes in distorted states [7] leading to static and dynamic cooperative Jahn-Teller effect in the boron sub-lattice of $RB_{12}$ crystals [6]. It was found in [7,8] that at low temperatures $T < T^* \sim 60$ K in the cage-glass disordered phase [9] the ferrodistortive effect on $B_{12}$ clusters causes the emergence of spatial modulation of the conduction electron density (dynamic charge stripes), which was confirmed by precise x-ray diffraction experiments on $LuB_{12}$, provoking also a strong anisotropy of magnetoresistance in combination with Hall effect singularities [10]. So far as the antiferromagnetic (AF) ground state in $RB_{12}$ with magnetic rare earth (RE) ions is formed by the indirect exchange interaction through conduction electrons (RKKY mechanism), it turns out that this is strongly affected by this (stripe) electron density redistribution. For example, in $Ho_xLu_{1-x}B_{12}$ antiferromagnets it leads to the formation of a complex *H-T-φ* magnetic phase diagram with numerous phases [11-13] and with very strong Maltese Cross type anisotropy. Taking into account that alike $HoB_{12}$ also $TmB_{12}$ shares the same *fcc* crystal structure with a Jahn-Teller instability of boron sub-lattice and in combination with a very similar magnetic structure and the same cubic single-ion anisotropy (triplet $\Gamma_{5(1)}$ ground state [14-16]), it is promising to investigate the effect of both the lattice instability and the dynamic charge stripes on the charge transport and magnetic ordering, and to compare these two non-equilibrium antiferromagnets.

## 2. Experimental details.

High quality single crystals were grown using crucible-free inductive floating zone melting in the inert gas atmosphere to prevent boron evaporation. Sample composition was verified by microprobe analysis, and structural quality and crystallographic orientation were checked by x-ray diffraction technique. Detailed information on the preparation of source $TmB_{12}$ sintered rods as well as details of the crystal growth are presented in [17, 18].

Specific heat (*C*) and magnetization (*M*) measurements were carried out on Quantum Design PPMS-9 and MPMS-5XL installations, correspondingly, both in the AF and in paramagnetic (P) states. MPMS SQUID-magnetometer was equipped with rotating sample insert, enabling angular dependencies to be obtained. The stationary sample mount setup provides high precision (~1%) absolute magnetization values which consider the demagnetizing factor, finite sample size and radial displacement inside magnetometer pickup coils. Charge transport experiments were conducted with an original setup [18], using standard 4-points direct current technique with circuit commutation. The axis of sample rotation in this precise angle-resolved resistivity (*ρ*) measurements was coincided with current direction $I \parallel [110]$ oriented transverse to external magnetic field $H \parallel (110)$. The studies have been carried out at temperatures in the range 1.8 – 300 K in magnetic field up to 80 kOe for resistivity, up to 90 kOe for specific heat and up to 50 kOe for magnetization.

## 3. Experimental results and data analysis.

*3.1 AF-P transition and charge transport anisotropy in the paramagnetic state.*

TmB$_{12}$ is antiferromagnetic metal (see fig.1a) with Neel temperature $T_N = 3.3$ K [19]. The phase transition at $T_N$ discerns clearly at $H = 0$ on the temperature dependencies of resistivity (Fig.1a), magnetic susceptibility (Fig.1b) and specific heat (Fig.1c). The AF state suppressed by external magnetic field is accompanied with emergence of (*i*) a strong (~35% in $H = 80$ kOe) negative magnetoresistance (MR) $\Delta\rho/\rho(H) = (\rho(H) - \rho(H=0))/\rho(H=0)$ (see fig.1a), (*ii*) a tendency to saturation of magnetization (decrease of low temperature magnetic susceptibility in strong magnetic field, see fig.1b) and (*iii*) magnetic Schottky anomaly in the heat capacity at temperatures in the range 2 - 20 K (fig.1c). The low field magnetic susceptibility $\chi(T)$ demonstrates Curie-Weiss type behavior

$$\chi = M/H = N_{Tm}\, \mu_{eff}^2 /(3k_B (T-\theta_p)) + \chi_0 \qquad (1),$$

where $N_{Tm} = 0.97 \cdot 10^{22}$ cm$^{-3}$ and $\mu_{eff} \approx 7.64\ \mu_B$ ($\mu_B$ and $k_B$ are the Bohr magneton and Boltzmann constant, correspondingly) are the concentration of Tm-ions and the effective magnetic moment per unit cell, $\theta_p \approx -17.2$ K is the paramagnetic Curie temperature corresponding to the AF exchange between these magnetic dipoles and $\chi_0 \approx -1.03 \cdot 10^{-7}\ \mu_B/\text{Oe}$ is the temperature independent term composed of the diamagnetic susceptibility of the boron cage and Pauli paramagnetism of conduction electrons. A fitting of the experimental curve $\chi(T)$ by Eq. (1) indicates that within the limits of experimental accuracy the susceptibility follows the Curie-Weiss dependence with about total Tm$^{3+}$ magnetic moment in the range 120 - 300 K. But, as the population of excited magnetic states of the crystal field splitting multiplet $^3H_6$ declines strongly below 120 K [15, 20] the magnetic moment decreases only slightly in the interval 25 - 120 K (fig.1d), and a ferromagnetic correlations occur in the range 3.5 - 25 K. As estimated from Eq. (1) the changes of effective moment $\mu_{eff}(T)$ are shown in fig.1d. Note, that below 25 K various types of short range order anomalies were observed previously in TmB$_{12}$ including a ferromagnetic phase transition detected at $T_C \approx 2.6$ K [14, 21, 22]. The anisotropy of the magnetic response which is detected in strong magnetic field (fig.1b, see also fig.4b) will be analyzed in more detail in the next sections.

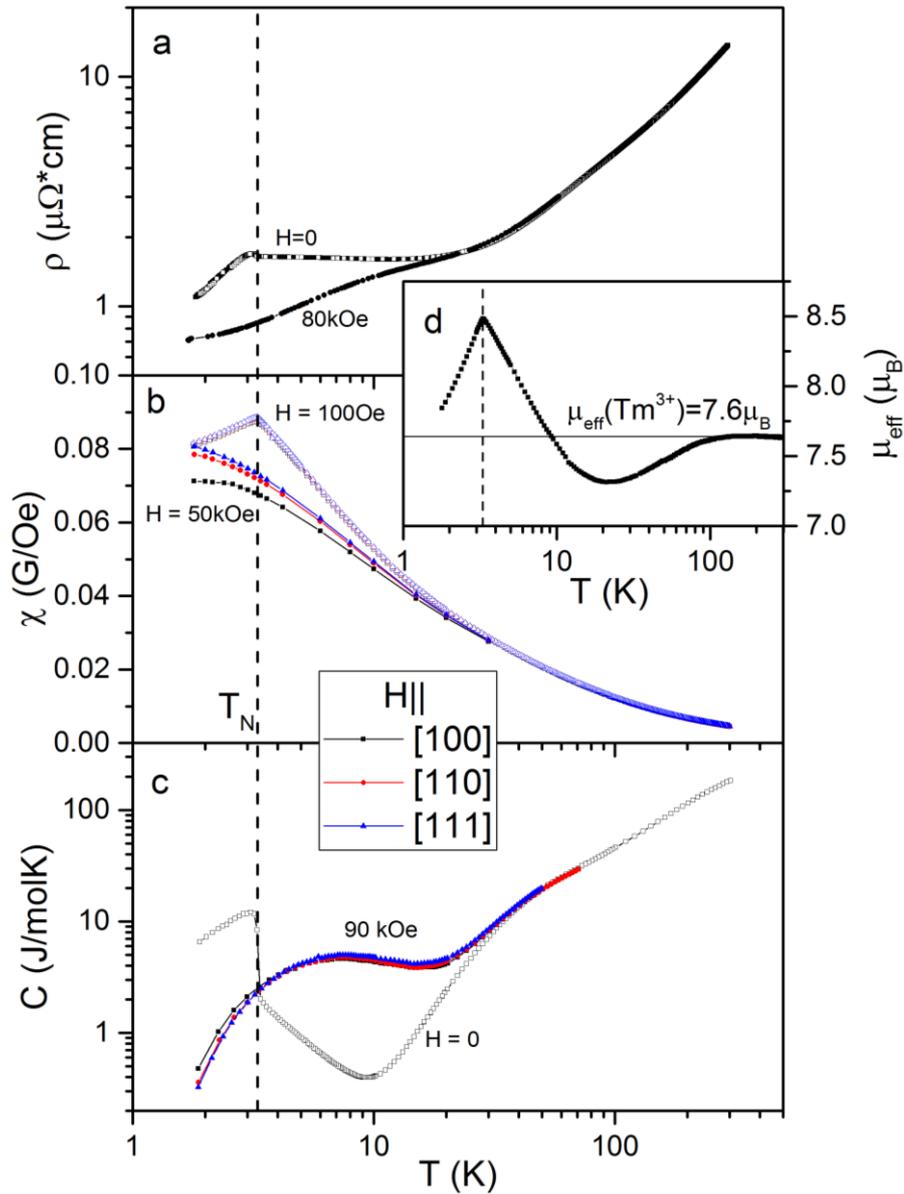

**Fig.1.** Temperature dependencies of (a) resistivity $\rho$, (b) magnetic susceptibility $\chi$ and (c) specific heat $C$ in magnetic field and for $H = 0$. (d) Effective magnetic moment as deduced from the Curie-Weiss analysis of equation (1).

Results of magnetoresistance $\Delta\rho/\rho(H)$ measurements in the external magnetic field up to 80 kOe are shown in fig.2. Both the field and angular scans at helium temperature T=4.2 K are presented, demonstrating quite different field dependences for various **H** directions (Fig.2a) and a rather strong high field MR anisotropy (Fig.2b) which is similar to that observed in the diamagnetic metallic reference compound $LuB_{12}$ [7, 8, 10] and in paramagnetic $Ho_xLu_{1-x}B_{12}$ solid solutions [11-13]. The anisotropy of the positive component of MR was attributed previously to the charge carriers scattering on dynamic charge stripes along <110> directions. Indeed, it was shown in [7, 8, 10-13, 23] that the interaction of dynamic charge stripes (quantum oscillations of the electron density with a frequency of ~ 240 GHz [24]) with strong tranverse magnetic field leads to resistivity enhancement for **H**||[001] even at intermediate temperatures. It is worth noting that sample rotation around **I**||[110] axis in the transverse configuration allows to vary the **H** vector in the (110) plane in between three principal crystallographic directions [001], [1-10] and [111] of the cubic structure. Thus, the performed measurements provide the total MR anisotropy in RE dodecaborides. It is seen from fig.2 that in paramagnetic state the MR is negative, and in ~~the~~ moderate magnetic fields a quadratic isotropic dependence $-\Delta\rho/\rho(H) \sim H^2$ is observed. On the contrary, in strong magnetic field an additional positive anisotropic component appears demonstrating itself with a prominent maximum on MR curves in vicinity of **H**||[001] direction. As a result, the angular dependence $\Delta\rho/\rho(\varphi)$ resembles the one found in diamagnetic reference compound $LuB_{12}$ [7, 8, 10] and seems to be alike to this observed in Ho-based magnetic dedecaborides [11-13].

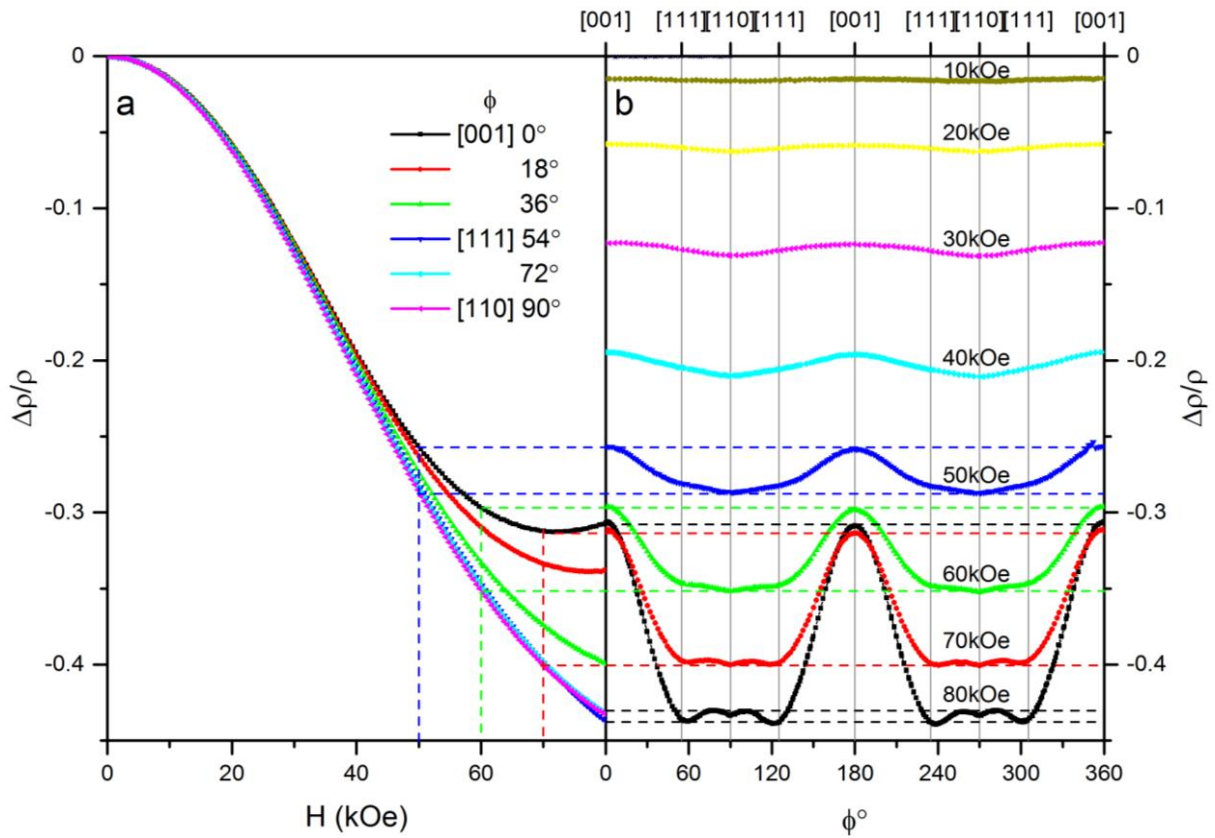

**Fig.2.** Field (a) and angular (b) dependencies of magnetoresistance $\Delta\rho/\rho(H, \varphi)$ in the paramagnetic phase at $T = 4.2$ K. Dash lines show the correspondence in amplitude between field and angular scans.

*3.2 H-T phase diagrams.*

When the temperature decreases below $T_N \approx 3.3$ K, in the AF state the magnetoresistance becomes alternating in sign and demonstrates multiple sharp anomalies, apparently corresponding to magnetic phase transitions (see fig.3a-3c). Although is the microscopic nature of these different magnetically ordered states yet to be revealed, one can reconstruct the corresponding phase diagram and conclude about charge carriers scattering regimes from the detailed analysis of magnetoresistance data. A set of magnetoresistance *vs* magnetic field curves $\Delta\rho/\rho(H)$ is presented on ~~the~~ panels of fig.3a - 3c and their derivatives $d(\Delta\rho/\rho)/dH(H)$ shown on figs.3d - 3f for orientations $H \parallel [001]$, $H \parallel [111]$ and $H \parallel [110]$, correspondingly, demonstrating singularities located at $H_{M1}$ - $H_{M3}$, $H_M$ and $H_N$ positions.

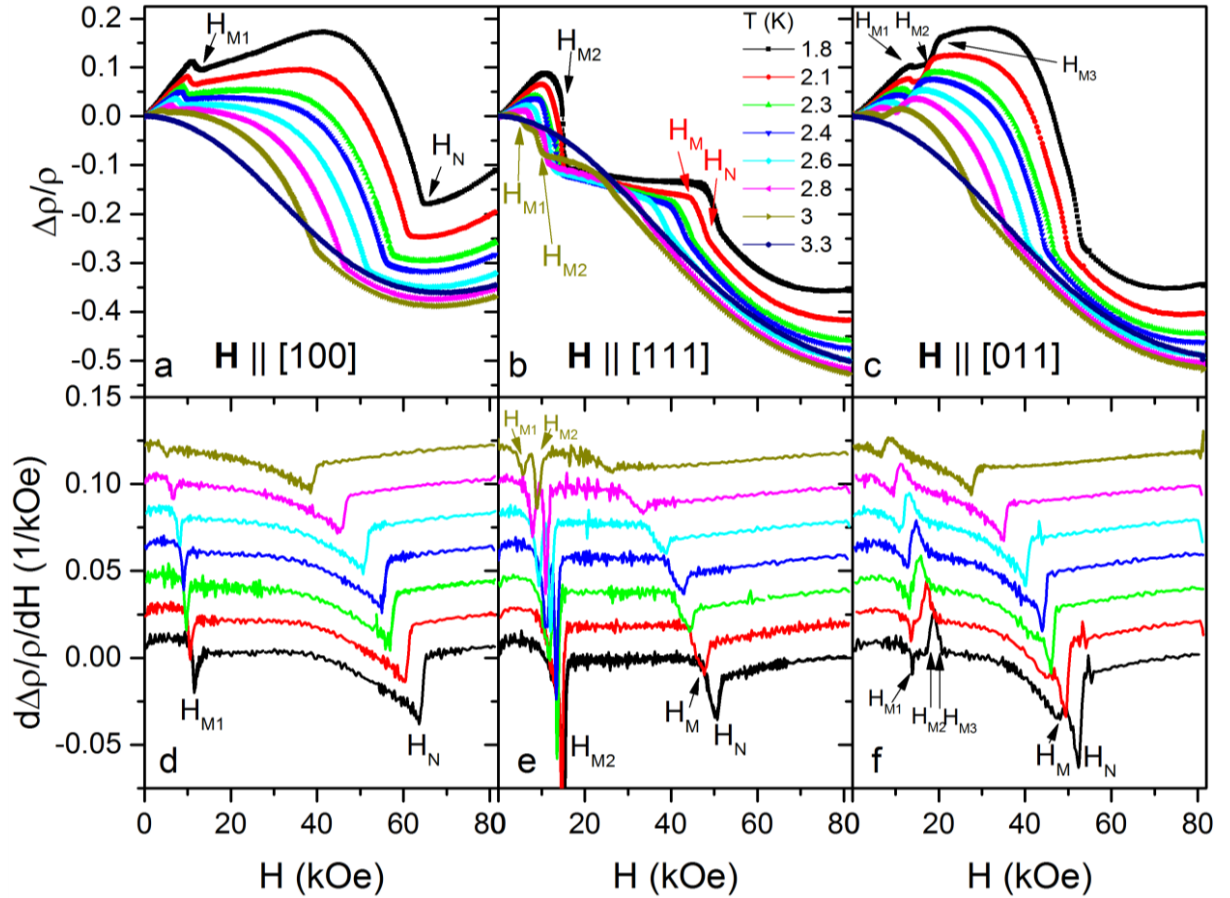

**Fig.3**. Field dependencies of magnetoresistance at different temperatures for $H \parallel [001]$ (a), $H \parallel [111]$ (b) and $H \parallel [011]$ (c), and their derivatives $d(\Delta\rho/\rho)/dH(H)$ (d),(e),(f) correspondingly. Derivative curves are shifted consistently upward to be visible.

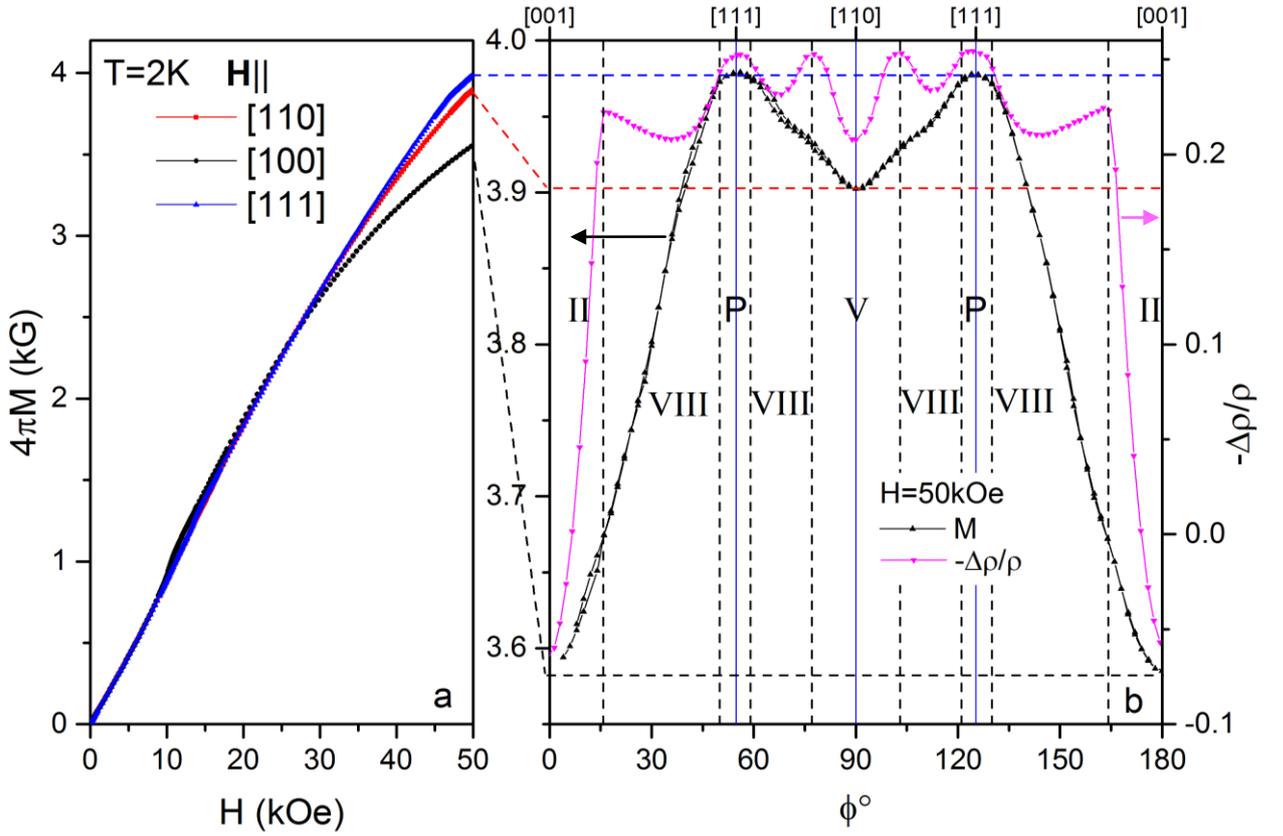

**Fig.4.** (a) Field dependencies of magnetization in AF-phase at $T = 2$ K for $H$ parallel to three principal crystallographic directions in the *fcc* lattice, and (b) comparison of enlarged scale angular dependencies of magnetization $M(\varphi)$ (black curve, left axis) and magnetoresistance $\Delta\rho/\rho(\varphi)$ (pink curve, right axis) at $H = 50$ kOe and $T = 2$ K. Roman numerals denote magnetic phases, P- paramagnetic state.

To refine the *H-T* magnetic phase diagram we have studied the field dependences of magnetization, field and temperature dependences of heat capacity and compared the data sets obtained for $H$ oriented along three principal crystallographic axes with the MR results of fig.3. An example of magnetization $M(H, T_0=1.8$ K$)$ field scans measured on three samples cut from one single crystal of $TmB_{12}$ is presented in fig.4a. It is seen from fig. 4 that in small fields H < 10 kOe in the AF state the magnetization is about linear and almost isotropic. When the magnetic field increases above 10 kOe several kinks are observed on *M(H)* curves which may be attributed to (*i*) orientation phase transitions in the AF phase and (*ii*) emergence of the spin-polarized paramagnetic (P) state in strong field. A full array of $M(H,T_0)$ curves for three principal directions is shown in [44] (see Fig.S1). High precision magnetization measurements allow us to fulfill the numerical differentiation of data obtained. The set of derivatives $dM/dH(H)=\chi(H, T_0)$ is shown in fig.5. Depending on $H$ direction the number of anomalies in the AF state changes from one for $H \parallel [001]$ (marked by $H_{M1}$ in fig.5a) to three for $H \parallel [011]$ and $H \parallel [111]$ ($H_M$, $H_{M1}$-

$H_{M3}$ in fig.3b and 3c). At high fields a step-like singularity of susceptibility is observed which corresponds to the location of the Neel field $H_N$ (the AF-P phase transition). As a result, for **H** || [100] direction we observe only one orientation transition at low magnetic field, and the location of AF-P phase boundary lies out of the available range of magnetization measurements ($H_N > 50$ kOe, fig.5a). On the contrary, both for **H** || [110] and **H** || [111] this step-like anomaly is clearly detected on $\chi(H, T_0)$ curves and can be attributed to the emergence of the field polarized P-state.

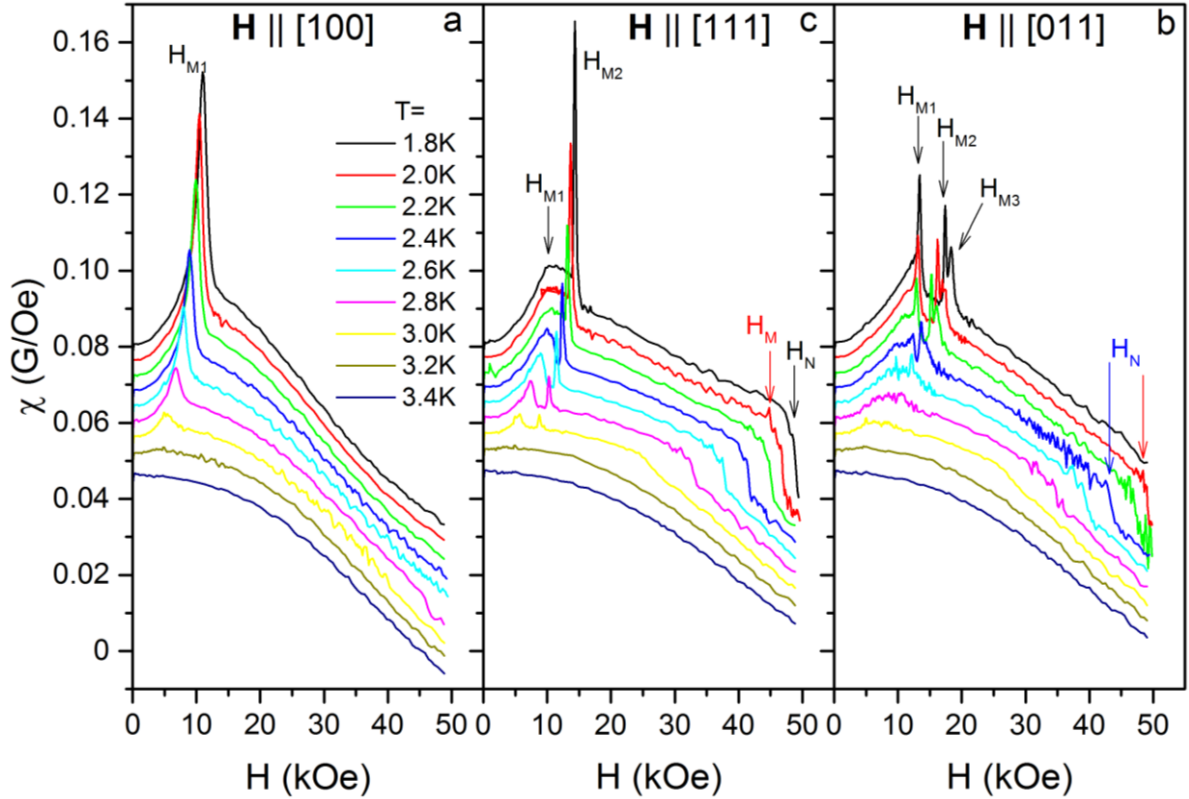

**Fig.5**. Magnetic susceptibility $\chi(H, T_0)$ for **H** || [100] (a), **H** || [110] (b) and **H** || [111] (c) at different temperatures. Curves are vertically shifted to be better distinguishable.

Note that for **H** || [111] the $\chi(H)$ curves demonstrate a very sharp peak preceded by broader maxima and small anomalies can be observed slightly below $H_N$ (Fig.5c). Refined *H-T* magnetic phase diagrams are shown in fig.6 combining the magnetization results with the data obtained from specific heat (see fig.S2 in supplementary information [44] for more details) and resistivity measurements (fig.3).

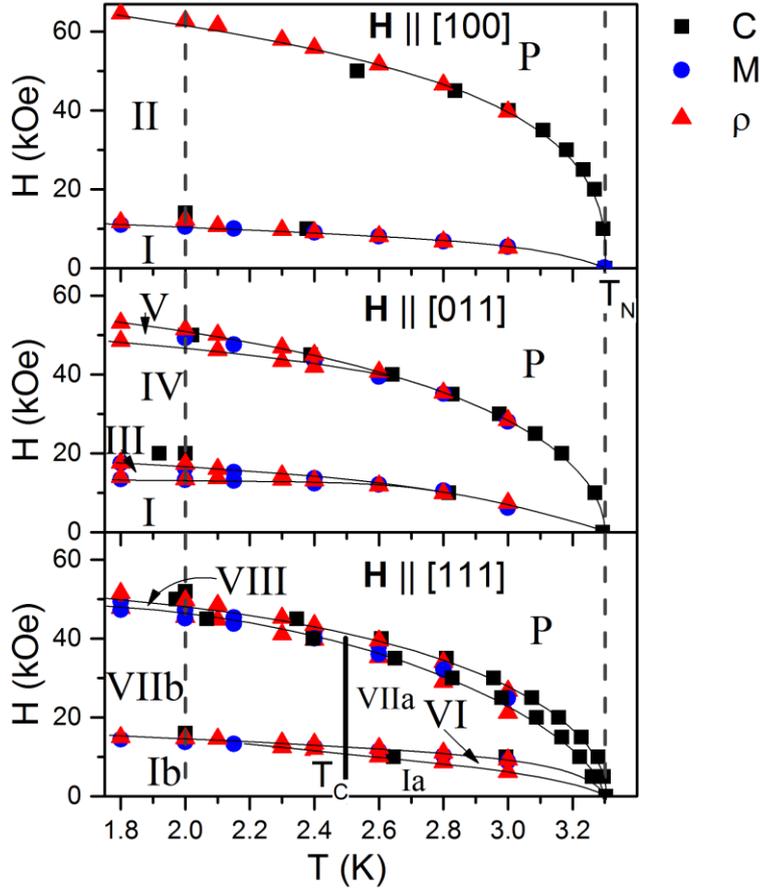

**Fig.6**. H-T phase diagram of $TmB_{12}$ for three principal directions.

It will be shown below that although the phase diagrams in panels (a)-(c) of fig.6 look quite similar, only the low field AF phase (marked as Ia in fig.6) and the high field spin-polarized P- phase are common for all three principal directions of magnetic field. On the contrary, the numerous magnetic phases II-X in fig.6 are peculiarly separated from each other both by radial and circular phase boundaries which characterize distinct regimes of charge carrier scattering. The features and regimes of the charge carrier scattering will be in more detail analyzed in the next sections.

*3.3 H-φ phase diagrams in the (110) plane at T = 2 K.*

To specify the location of the phase boundaries for various strength and orientation of **H** situated in the (110) plane we have measured at $T = 2$ K the angular dependences of magnetization $M(\varphi, H_0)$ and precise angle-resolved $\Delta\rho/\rho(\varphi, H_0)$ at fixed magnetic fields (see Figs.7a-7g and 7f, correspondingly, and fig.4b) from one side, and complementary field dependences of these two characteristics at fixed angles $\angle\varphi = [001]\wedge H$ in the range 0-90°. Fig.8a shows the performed magnetic field derivatives $dM/dH(H)=\chi(H, \varphi_0)$, the observed MR changes are shown in figs. 8b and 8c. It is seen from fig.8 that the single feature at $H_{M1}$ observed for $\varphi =$ 0 ($H \parallel [001]$) (see also fig.5a) transforms into three anomalies at $H_{M1}$, $H_{M2}$ and $H_{M3}$ when $\varphi$ increases and passes through $\varphi = 54.7°$ ($H \parallel [111]$) up to 90° ($H \parallel [110]$). Below $H_N$ an additional anomaly is observed (see also MR derivatives in Fig.S3 in [44]). Similar to magnetization (Fig.4a) the MR in the AF state is about isotropic in the low field interval $H<10$ kOe when $\Delta\rho/\rho$ is positive and has an about linear magnetic field dependence.

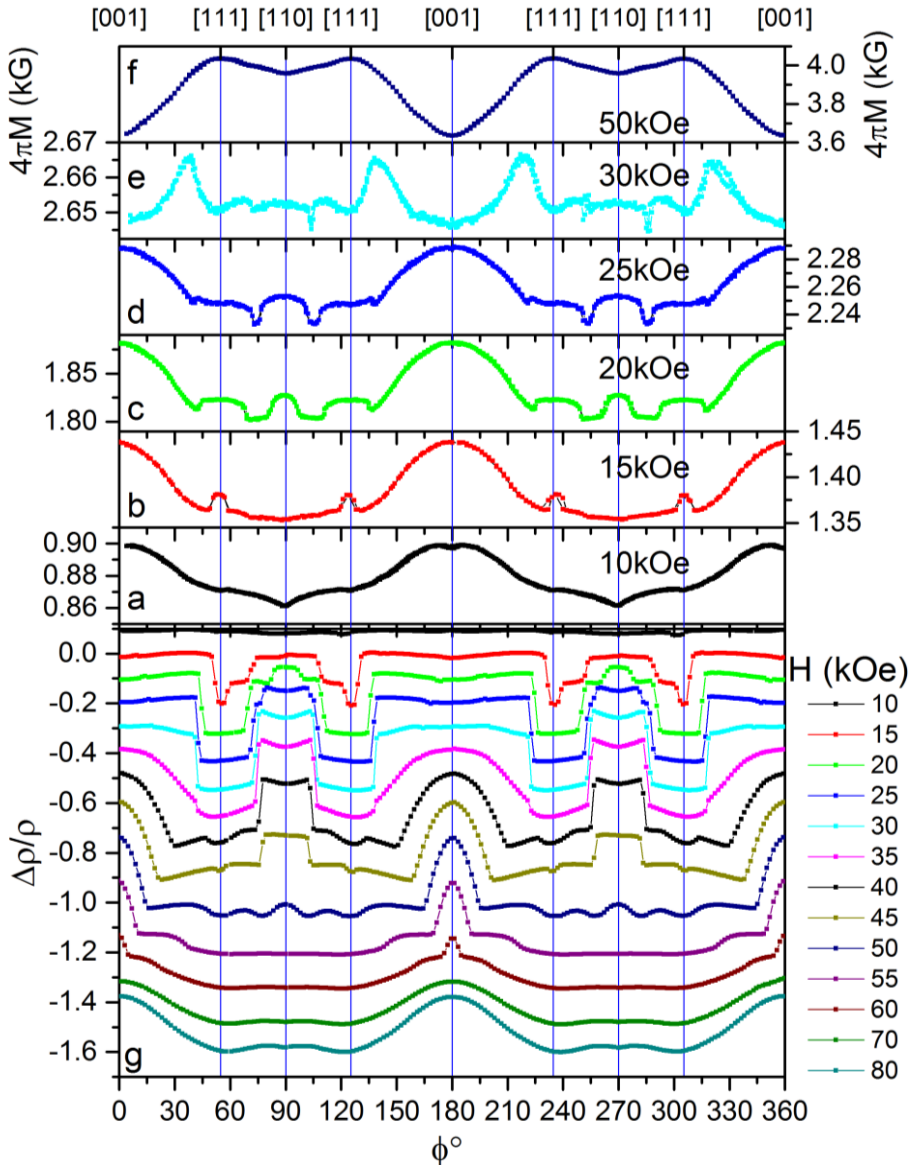

**Fig.7**. Angular dependencies of TmB$_{12}$ (a-f) magnetization and (g) magnetoresistance for the various magnetic fields at $T = 2$ K. The rotation was performed around the axis $I \parallel [011]$. Vertical lines show positions when the magnetic field $H$ is aligned with the principal axes. Curves on (g) are shifted for convenience.

On the contrary, above 10 kOe the MR becomes strongly anisotropic and one can distinguish three main angular segments $\Delta\varphi_{001}$, $\Delta\varphi_{110}$ and $\Delta\varphi_{111}$ (see figs.8b-8c), which are located around three principal crystallographic directions and separated one from another by abrupt radial and circular borders. Firstly, the MR anisotropy appears mainly due to the large magnitude anomaly at $H_{M1}$ and then, either an about linear increase of $\Delta\rho/\rho(H, \varphi_0)$ (in $\Delta\varphi_{001}$), or, decrease (in $\Delta\varphi_{111}$) is observed depending from the sector examined (see fig.8b and 8c). It is worth noting that the location of AF-P transition at $H_N(\varphi)$ demonstrates at $T = 2$ K a perceptible anisotropy change from 49 kOe for **H** ∥ [111] to 62 kOe in the orientation **H** ∥ [001] (see, for example the dashed line on the right in fig.8b). Note also, that when **H** direction changes inside the $\Delta\varphi_{001}$ and $\Delta\varphi_{111}$ sectors the phase transition at $H_N$ splits in two, and the lower singularity shifts to lower fields. As these high field anomalies move apart, linear MR range shows in between with negative slope growing in absolute value. We found also some small but discernible MR features in fields 16 kOe ≤ $H$ ≤ 26 kOe and 4.5° ≤ $\varphi$ ≤ 20°, whose location has uncertainty due to hysteresis.

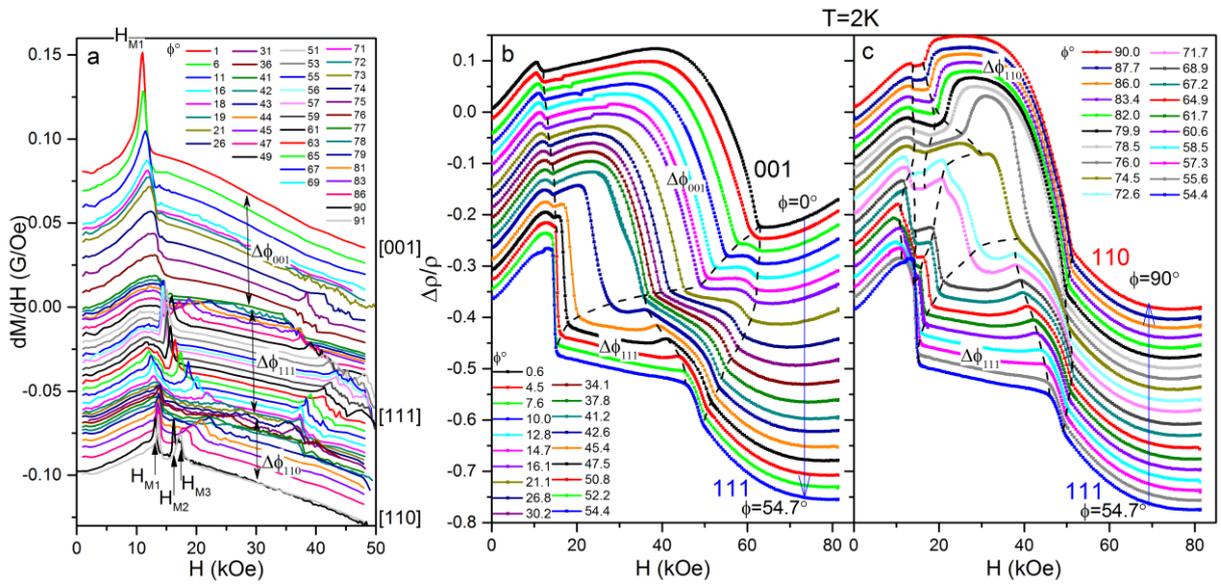

Fig.8. Magnetic field dependencies of TmB$_{12}$ - (a) magnetization derivatives and (b)-(c) magnetoresistance at $T = 2$ K for different angles ∠$\varphi$ = [001]^**H** between the [001] crystallographic direction and magnetic field **H**. $H_{M1} - H_{M3}$ marked with arrows on (a) and the dash lines on (b)-(c) show the orientational magnetic phase transitions. Angular ranges $\Delta\varphi_{001}$, $\Delta\varphi_{110}$, $\Delta\varphi_{111}$ are discussed in text. The rotation was performed around the axis **I** ∥ [110]. Curves are shifted for convenience (proportional to the angle $\varphi$ in (a), consistently downward on (b) and consistently upward on (c)).

The results of angle-resolved magnetization $M(\varphi, H_0)$ and precise ARMR $\Delta\rho/\rho(\varphi, H_0)$ measurements at fixed magnetic fields shown in fig.7 elaborate the picture of phase transitions in the ($H, \varphi$) plane drawn in field sweep experiments (Fig.8). In particular, the $\Delta\rho/\rho(\varphi, H_0)$ behavior displays a very small anisotropy below 10 kOe, emerging step-like borders between $\Delta\varphi_{100}$, $\Delta\varphi_{111}$ and $\Delta\varphi_{110}$ segments and their transformation when $H$ increases in the range $H_{M1} < H$

< $H_N$. Note, that the anomalies on the angular $M(\varphi, H_0)$ and $\Delta\rho/\rho(\varphi, H_0)$ curves appear at the same directions of **H** (see, for example, the comparison presented in fig. 4b) demonstrating synchronous changes in magnetic and charge transport characteristics attributed to orientation phase transitions.

An overall view of the results of $\Delta\rho/\rho(H,\varphi)$ measurements at $T = 2$ K presented in figs.7b and 8b-8c is displayed in fig. 9a in the cylindrical coordinates where the MR is plotted along the vertical axis and is additionally provided with a color scale. For comparison a similar view of the MR distribution in the paramagnetic state at $T = 4.2$ K is shown in fig.9c and the projections of these two data sets onto $(1\bar{1}0)$ plane (see figs.9b and 9d) allow us to refine the $(H,\varphi)$ AF phase diagram ($T = 2$ K, panel b) and to link its symmetry with the anisotropy of charge carriers scattering in the P-phase ($T = 4.2$ K, panel d). Roman numerals in panel (b) show different magnetic I-X phases in the AF state (see Fig.S5 in [44] for more details).

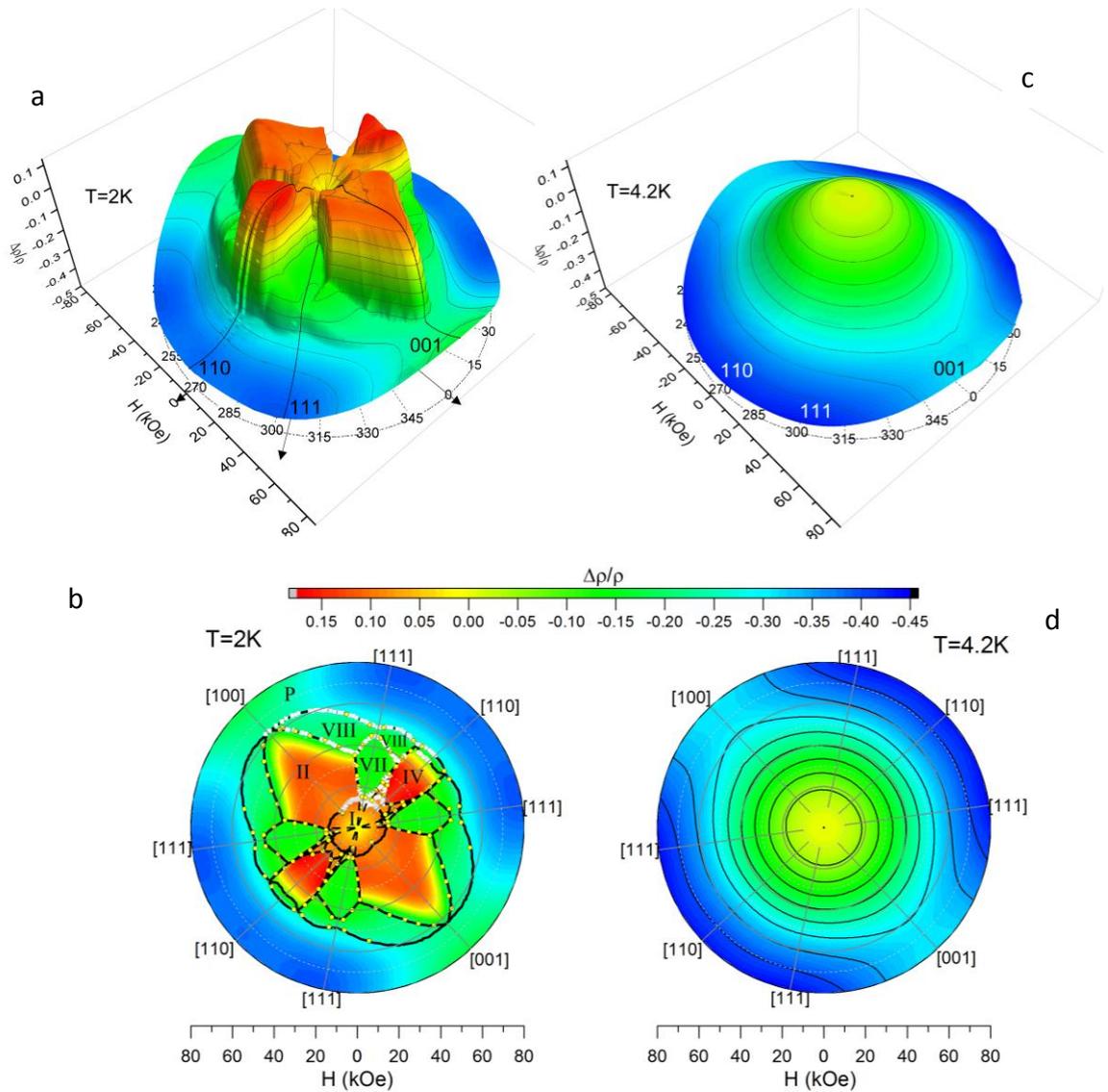

**Fig.9.** Magnetoresistance $\Delta\rho/\rho = f(H,\varphi)$ of TmB$_{12}$ in cylindrical coordinates for current directions **I** ∥ [1$\bar{1}$0] at (a) T = 2 K (AF phase) and (c) T = 4.2 K (P phase) and its projection onto (1$\bar{1}$0) planes (b) and (d) respectively (see [44] for more details). Roman numerals in panel (b) show different magnetic phases in the AF state. White points were taken from field scans, yellow ones from angular scans.

It can be seen from fig.9b that despite high symmetry of the crystal lattice *the magnetic phases for the principal field directions* are separated by the radial and circular phase boundaries and, hence, these phases *are completely different,* and that only one low-field AF state (marked as Ia in fig.6) exists for any magnetic field orientation. The phase boundaries on *H-φ* diagrams in the (110) plane (symbols in figs. 6 and 9b are related to sharp features detected on the $\Delta\rho/\rho$ angular dependencies, on derivatives of MR and magnetization (Fig.8a and Fig.S3 in [44]). Several regions corresponding to the phases with different magnetically ordered states are easily resolved in fig.9b. Similar "Maltese Cross" anisotropy in the (1$\bar{1}$0) plane was found recently in related Ho$_x$Lu$_{1-x}$B$_{12}$ and Tm$_{1-x}$Yb$_x$B$_{12}$ systems [11-13, 25].

## 4. Discussion

Firstly, we need to mention the similarity of the *H-T* and *H-φ* phase diagrams of TmB$_{12}$ and HoB$_{12}$ which are characterized by the same cubic single-ion anisotropy ($\Gamma_{5(1)}$ ground state of the 4f shell) and the alike set of conduction band parameters [26]. When comparing the Maltese cross type phase diagrams of HoB$_{12}$ [13] and TmB$_{12}$ (Fig.9b), note that in the case of TmB$_{12}$ a considerable Neel field anisotropy (more than 25 % at *T* = 2 K) appears additionally with new phases on the *H-φ* diagram in the (110) plane (Fig.9b). Moreover, phase I is no longer a common for any direction of magnetic field in TmB$_{12}$ (see fig.9b and fig.S5 in [44]). The combination of negative magnetoresistance contribution which is quadratic in moderate magnetic field with an about linear positive MR component (see fig.2 and figs.8b-8c) is also very similar in TmB$_{12}$ to that one found previously in the case of HoB$_{12}$ [27]. Strong and unusual effect in magnetoresistance of TmB$_{12}$ is the emergence of negative linear MR term in phase VII detected in the sector $\Delta\varphi_{111}$ (see Fig.9b).

A positive linear MR term deduced in the AF state is traditionally associated with charge carrier scattering on spin and charge density waves (SDWs and CDWs) (see, e.g., Refs. [21, 28, 29]). Such behavior is typical, for example, for the itinerant AF phase of chromium, in the absence of Fe and Co magnetic impurities [30] and it has been reported in systems with CDW order including 2H-NbSe2 [31], Nb3Te4 [32], (PO2)4(WO3)8 [33], and RTe3 (R = Tb and Ho)

[34]. The positive linear MR in the low-field limit has been attributed to a wide range of mechanisms, from magnetic breakdown [31] to scattering on CDW fluctuations at Fermi surface hot spots [34]. Recently the authors [35] have argued that there exists a universal mechanism to create linear positive MR in the low-field/low-temperature limit which is well suited for explaining the behavior of a broad family of density wave materials in which large patches of the Fermi surface are gapped out by the formation of long-range order. The salient features are a consequence of itinerant carriers turning sharp corners of the Fermi surface. This mechanism is greatly enhanced in both CDW and SDW systems, as the formation of correlated electronic states opens a gap at the Fermi surface, removing sheets of open electron paths while keeping only small electron/hole pockets/ellipsoids with small orbits of sharp curvature. Although only representing a small portion of the Fermi surface, they can manifestly dominate the response to an applied field [35].

On the other hand, the negative quadratic MR component in non- Kondo-type systems with metallic conductivity and localized magnetic moments is attributed usually to electron scattering (*i*) in a paramagnetic state on a nanometer size clusters of magnetic ions, or (*ii*) on the on-site *4f-5d* spin fluctuations in the AF phase [11, 36]. The above mechanisms present in the AF and P states of $TmB_{12}$ are accompanied with charge carrier scattering on dynamic charge stripes. Following to [11, 36] we conclude that these fluctuating charges are among the main factors that determine the complexity of the magnetic phase diagram in this strongly correlated electron system. This kind of electronic instability is accompanied with nanometer scale phase separation in combination with local *4f*-*5d* fluctuations of electron density in the nearest vicinity of $Tm^{3+}$ ions. The positional disorder in the arrangement of $Tm^{3+}$ ions which are loosely bound in $B_{24}$ truncated cubooctahedrons leads in the cage-glass state at $T < T^* \sim 60$ K to a significant dispersion of exchange constants (through indirect exchange, RKKY mechanism) and formation of both nanometer size clusters of magnetic $Tm^{3+}$ ions in the $RB_{12}$ matrix (short-range order effects well above $T_N$) and to creation of strong local *4f-5d* spin fluctuations responsible for the polarization of *5d* conduction band states (the spin-polaron effect). The last one produces spin polarization subnanometer size ferromagnetic domains (ferrons, in the terminology of Refs. [37, 38]) resulting in stabilization of SDW antinodes in the $TmB_{12}$ matrix (see the increase of $\mu_{eff}$ below 25 K in the inset in fig.1b). The spin-polarized *5d* component of the magnetic structure (ferrons) assembled in SDW is from one side very sensitive to the external magnetic field [39, 40], and, from another side, the applied field suppresses *4f* -*5d* spin fluctuations by destroying the spin-flip scattering process. Moreover, along the direction of the dynamic charge stripes [110] in the *fcc* lattice the huge charge carrier scattering destroys the indirect exchange of the nearest neighbored $Tm^{3+}$ localized magnetic moments located at the distance of 5.3 Å from each other, renormalizing the RKKY interaction and accumulating a noticeable part of charge carriers into a filamentary electronic structure. Thus, the complex $H$-$\varphi$ phase diagrams of TmB$_{12}$ antiferromagnet may be explained in terms of the formation of *a composite magnetically ordered state of localized 4f magnetic moments of Tm$^{3+}$ ions in combination with spin polarized local areas of 5d states (ferrons) involved in the formation of SDW* in the presence of a filamentary structure of dynamic charge stripes [11, 36].

Then, turning to the additional negative linear magnetoresistance components appeared solely in phase VII (sector $\Delta\varphi_{111}$, see fig.9b) it is worth noting that this kind of behavior is typical for metallic ferromagnets below Currie temperature, as the magnetic field quenches spin fluctuations and reduces spin scattering of itinerant carriers [41, 42] (see, for example, the magnetoresistance of Ni [43]. From this point of view, in TmB$_{12}$ the emergence of the negative linear MR term may be attributed to the strong scattering on the spin-polarized (ferron) component of the magnetic structure which is quite different from that one observed in HoB$_{12}$ [13]. According to estimations of [42] the relation

$$-\Delta\rho/\rho(H) \sim H/(1-T/T_C)^{1/2} \qquad (2)$$

($T_C$ is the Currie temperature of ferromagnetic nanodomains) should be valid in this regime, and the linear negative MR term demonstrates an appropriate scaling for TmB$_{12}$ in $\Delta\varphi_{111}$ sector below $T_C \approx 2.5$ K $< T_N$ (see fig.S4 in [44]). Such features at $T_C$ were found in this study also on resistivity derivatives $d\rho/dT(T)$ (fig.S4a in [44]) and detected clearly for the Mott thermopower product $S/T=f(T)$ (see fig.S4c in [44], $S$- Seebeck coefficient) which was calculated from the data of [45]. Finaly, a detailed quantitative analysis of the MR components in three magnetic dodecaborides REB$_{12}$ (RE - Ho, Er and Tm) is in progress and the results will be bulished elsewhere.

**Conclusion**

To summarize, the model strongly correlated electron system TmB$_{12}$ with a *fcc* lattice instability (cooperative dynamic JT effect) and nanometer size electronic phase separation (dynamic charge stripes) has been studied in detail by low temperature magnetoresistance, heat capacity, and magnetization measurements. The $H$-$T$ and angular $H$-$\varphi$ magnetic phase diagrams in the form of a Maltese cross with multiple magnetic phases and a notably anisotropic Neel field $H_N(\varphi)$ have been deduced for this AF metal, and matched to three main sectors in vicinity of the main directions (*i*) ***H*** || [110] along dynamic charge stripes, (*ii*) ***H*** || [001] transverse to dynamic charge stripes, and (*iii*) ***H*** || [111] connected with the orientation of the AF magnetic structure. A Maltese cross type anisotropy was detected for charge carriers scattering, and three dominated mechanisms of magnetoresistance in the complicated AF state were discussed being attributed to charge carrier scattering on (*i*) SDW (*5d* component of magnetic structure), (*ii*) local *4f-5d* spin fluctuations of $Tm^{3+}$ ions and (*iii*) itinerant ferromagnetic nano-domains composed from polarized conduction electrons. We argue that similar to the case of $HoB_{12}$ the observed dramatic symmetry lowering in $TmB_{12}$ is the consequence of strong renormalization of the indirect exchange interaction (RKKY mechanism) due to the presence of dynamic charge stripes in the matrix of this AF metal.

**Acknowledgments**

This work was supported by the Russian Science Foundation, Project No. 21-12-00186 and was performed using the equipment of the Shared Facility Center for Studies of HTSC and Other Strongly Correlated Materials, Lebedev Physical Institute, the Russian Academy of Sciences, and of the Center of Excellence, Slovak Academy of Sciences. The work of K.F. and S.G. is supported by the Slovak agencies VEGA (Grant No. 2/0032/20) and APVV (Grant No. 17–0020). The work of A.A. and A.B. is supported by European Microkelvin Platform (Projects №21 and 22).

**References**

[1] Flachbart K, Alekseev P, Grechnev G 2008 Rare earth dodecaborides-magnetism, superconductivity and other properties *Rare Earths: Research and Applications* ed Delfrey K N (New York: Nova Science) p 79

[2] Mori T 2008 Higher borides *Handbook on the Physics and Chemistry of Rare Earths* vol 38 ed Gschneidner Jr. K A, Bunzli J-C G, Pecharsky V K (Amsterdam: Elsevier) p 105

[3] Sluchanko N E 2021 Magnetism, Quantum Criticality, and Metal–Insulator Transitions in $RB_{12}$ *Rare-Earth Borides* ed Inosov D S (Singapore: Jenny Stanford Publishing), Chapt.4

[4] Gabani S, Flachbart K, Siemensmeyer K, Mori T 2020 *J. Alloys Compounds.* **821** 153201

[5] Azarevich A N et al 2021 *Phys Rev B* **103** 104515

[6] Bolotina N B, Dudka A P, Khrykina O N, Mironov V S 2021 Crystal structures of dodecaborides: complexity in simplicity *Rare-Earth Borides* ed Inosov D S (Singapore: Jenny Stanford Publishing), Chapt. 3

[7] Sluchanko N et al 2018 *Phys Rev B* **97** 035150

[8] Bolotina N B, Dudka A P, Khrykina O N, Krasnorussky V N, Shitsevalova N Yu, Filipov V B, Sluchanko N E 2018 *J. Phys.: Condens. Matter* **30** 265402

[9] Sluchanko N E et al 2011 *J. Exp. Theor. Phys.* **113** 468

[10] Sluchanko N, Azarevich A, Bogach A, Demishev S, Krasikov K, Voronov V, Filipov V, Shitsevalova N, Glushkov V 2021 *Phys Rev B* **103** 035117

[11] Khoroshilov A L et al 2019 *Phys Rev B* **99** 174430

[12] Sluchanko N et al 2020 *Acta Phys. Polon. A* **137** 756

[13] Krasikov K M et al 2020 *Phys Rev B* **102** 214435

[14] Siemensmeyer K, Habicht K, Lonkai Th, Mat'aš S, Gabáni S, Shitsevalova N, Wulf E, Flachbart K 2007 *J Low Temp Phys* **146** 581

[15] Czopnik A, Murasik A, Keller L, Shitstevalova N, Paderno Y 2000 *Phys. Stat. Sol. b* **221** R7

[16] Siemensmeyer K, Flachbart K, Gabáni S, Maťaš S, Paderno Y, Shitsevalova N. 2006 *J. Solid State Chem.* **179** 2748

[17] Shitsevalova N Yu 2001 *PhD Thesis* Institute of Low Temperature and Structure Research, Wrocław (in Russian)

[18] Sluchanko N E, Azarevich A N, Bogach A V, Glushkov V V, Demishev S V, Anisimov M A, Levchenko A V, Filipov V B, Shitsevalova N Yu 2012 *J. Exp. Theor. Phys.* **115** 509

[19] Gabáni S, Baťko I, Flachbart K, Herrmannsdörfer T, König R, Paderno Y, Shitsevalova N 1999 *J. Magn. Magn. Mat.* **207** 131

[20] Alekseev P A, Nemkovski K S, Mignot J-M, Clementyev E S, Ivanov A S, Rols S, Bewley R I, Filipov V B, Shitsevalova N Yu 2014 *Phys Rev B* **89** 115121

[21] Sluchanko N E, Bogach A V, Glushkov V V, Demishev S V, Samarin N A, Sluchanko D N, Dukhnenko A V, Levchenko A V 2009 *J. Exp. Theor. Phys.* **108** 668

[22] Kalvius G M, Noakes D R, Marcano N, Wappling R, Iga F, Takabatake T 2003 *Physica B* **326** 398

[23] Sluchanko N E, Khoroshilov A L, Bogach A V, Voronov V V, Glushkov V V, Demishev S V, Krasnorussky V N, Krasikov K M, Shitsevalova N Yu, Filipov V B 2018 *JETP Letters* **107** 30

[24] Sluchanko N E et al 2019 *J. Phys.: Condens. Matter* **31** 065604

[25] Azarevich A et al 2020 *Acta Phys. Polon. A* **137** 788

[26] Pluzhnikov V, Shitsevalova N, Dukhnenko A, Czopnik A, Nizhankovskii V, Settai R, Ōnuki Y 2008 *J. Magn. Magn. Mat.* **320** 1597

[27] Krasikov K M et al 2020 *Solid State Sci.* **104** 106253

[28] Anisimov M A, Bogach A V, Glushkov V V, Demishev S V, Samarin N A, Filipov V B, Shitsevalova N Y, Kuznetsov A V, Sluchanko N E 2009 *J. Exp. Theor. Phys* **109** 815

[29] Anisimov M A, Bogach A V, Glushkov V V, Demishev S V, Samarin N A, Shitsevalova N Yu, Levchenko A V, Filipov V B, Sluchanko N E 2012 *J. Phys.: Conf. Ser.* **400** 032003

[30] Arajs S, Dunmyre G R 1965 *J. Appl. Phys.* **36** 3555

[31] Naito M, Tanaka S 1982 *J Phys Soc Jpn* **51** 228

[32] Ishihara Y, Nakada I 1983 *Solid State Commun* **45** 129

[33] Rötger A, Lehmann J, Schlenker C, Dumas J, Marcus J, Teweldemedhin Z S, Greenblatt M 1994 *Europhys Lett* **25** 23

[34] Sinchenko A A, Grigoriev P D, Lejay P, Monceau P 2017 *Phys Rev B* **96** 245129

[35] Yejun Feng et al 2019 *PNAS* **116** 11201

[36] Sluchanko N E et al 2015 *Phys Rev B* **91** 235104

[37] Nagaev E L 1967 *JETP Lett* **6** 18

[38] Kagan M Yu, Kugel K I, Khomskii D I 2001 *JETP* **93** 415

[39] Sasaki T, Lebed A G, Fukase T, Toyota N 1996 *Phys Rev B* **54** 12969

[40] Montambaux G 1988 Phys Rev B **38** 4788

[41] Yosida K 1957 *Phys Rev* **107** 396

[42] Yamada H and Takada S 1973 *J Phys Soc Jpn* **34** 51

[43] Pippard A B 1989 *Magnetoresistance in Metals* (Cambridge, UK: Cambridge Univ Press) pp 35–39

[44] Supplementary Information

[45] Sluchanko N E et al 2006 *Phys. Stat. Sol. b* **243** R63

# Supplementary information to the article
# Evidence of symmetry lowering in the antiferromagnetic metal TmB$_{12}$
# with dynamic charge stripes

A. Azarevich[1,2], V. Glushkov[1], S. Demishev[1], A. Bogach[1], V. Voronov[1], S. Gavrilkin[3], N. Shitsevalova[4], V. Filipov[4], S. Gabáni[5], J. Kačmarčík[5], K. Flachbart[5], N. Sluchanko[1]

[1]Prokhorov General Physics Institute of Russian Academy of Sciences, 38 Vavilov str., Moscow 119991, Russia

[2]Moscow Institute of Physics and Technology, 9 Institutskiy per., Dolgoprudny, Moscow Region, 141701, Russia

[3]Lebedev Physical Institute of RAS, 53 Leninsky Avenue, 119991 Moscow, Russia

[4]Frantsevich Institute for Problems of Materials Science, National Academy of Sciences of Ukraine, 3 Krzhizhanovsky str., Kyiv 03680, Ukraine

[5]Institute of Experimental Physics, Slovak Academy of Sciences, 47 Watsonova, 04001 Košice, Slovakia

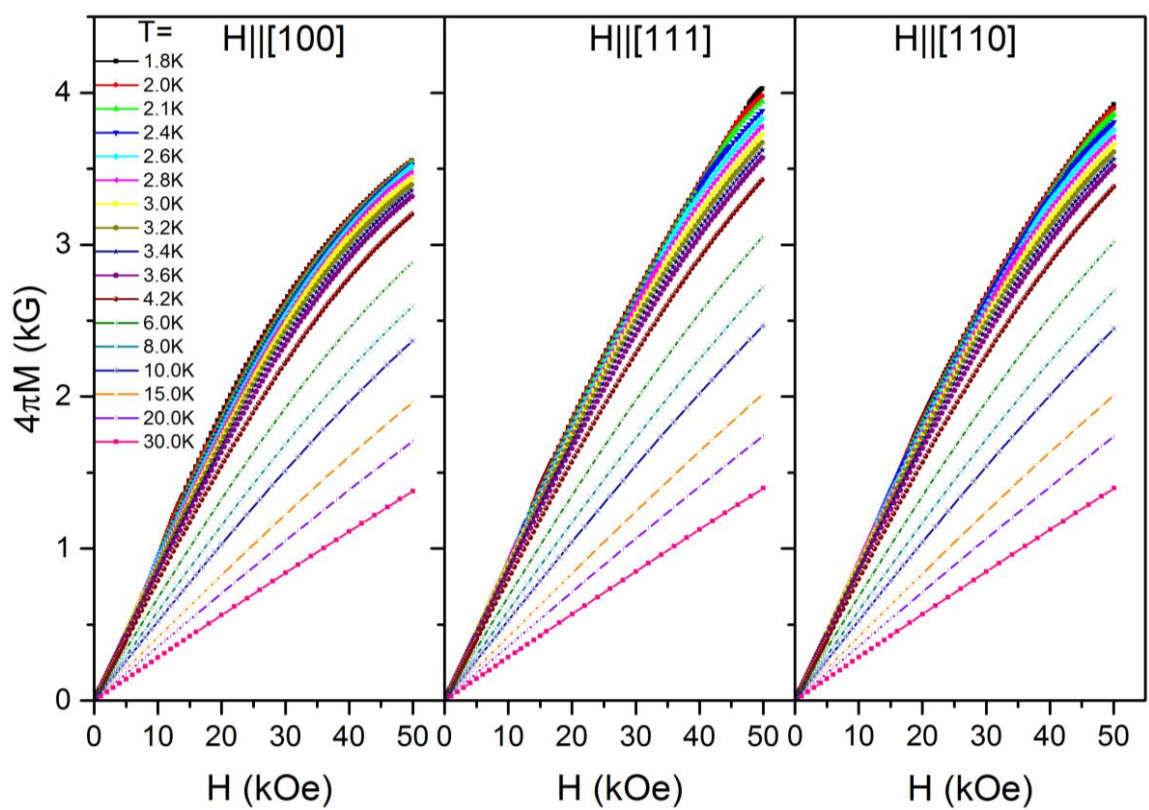

**Fig.S1** Magnetization as a function of external magnetic field, directed along tree principal crystallographic directions at different temperatures.

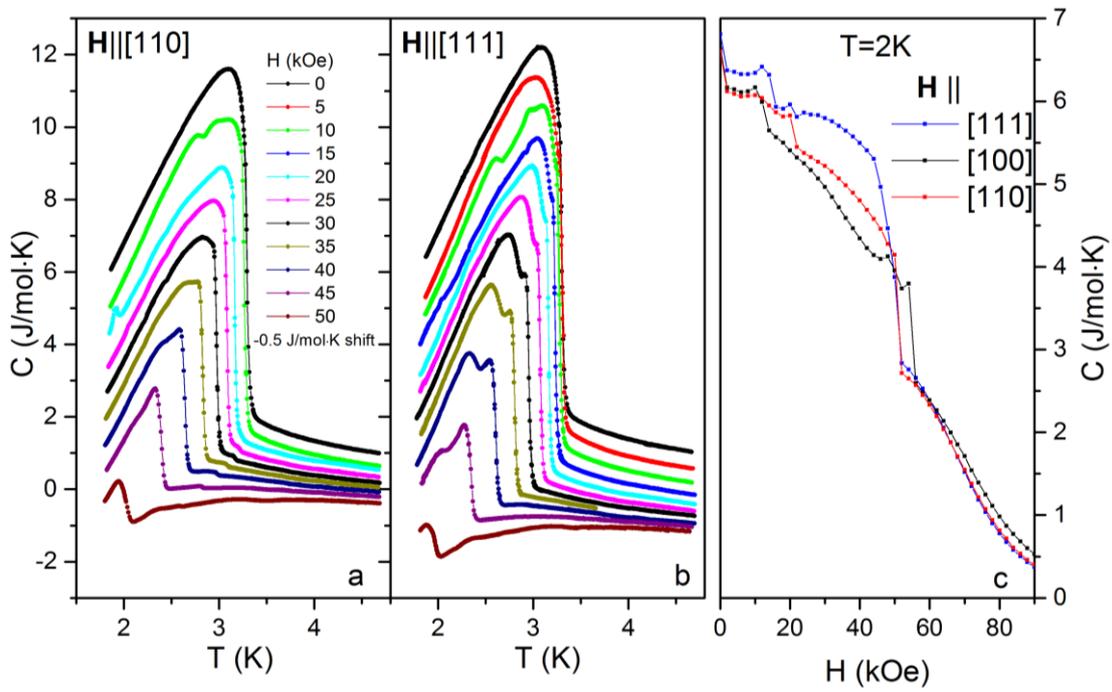

**Fig.S2** Specific heat C(T) in various magnetic field for (a) **H** ∥ [110] and (b) **H** ∥ [110], curves are shifted downward for convenience. (c) Field dependences C(H, $T_0$=2K) for **H** along three principal crystallographic directions.

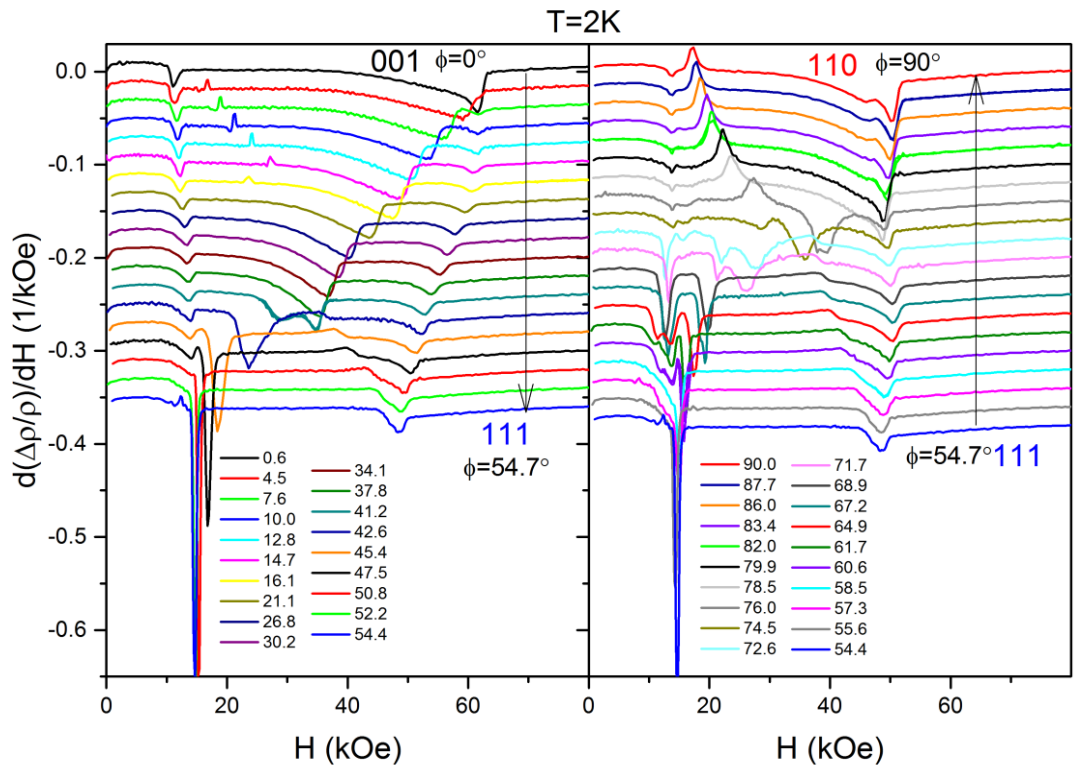

**Fig.S3** Magnetic field dependencies of magnetoresistance derivatives d(Δρ/ρ)/dH at $T$ = 2 K for different angles ∠φ =**n^H** between the normal **n** ∥ [001] direction in the crystal and magnetic field **H**. The rotation was performed around the axis **I** ∥ [110]. Curves are shifted for convenience.

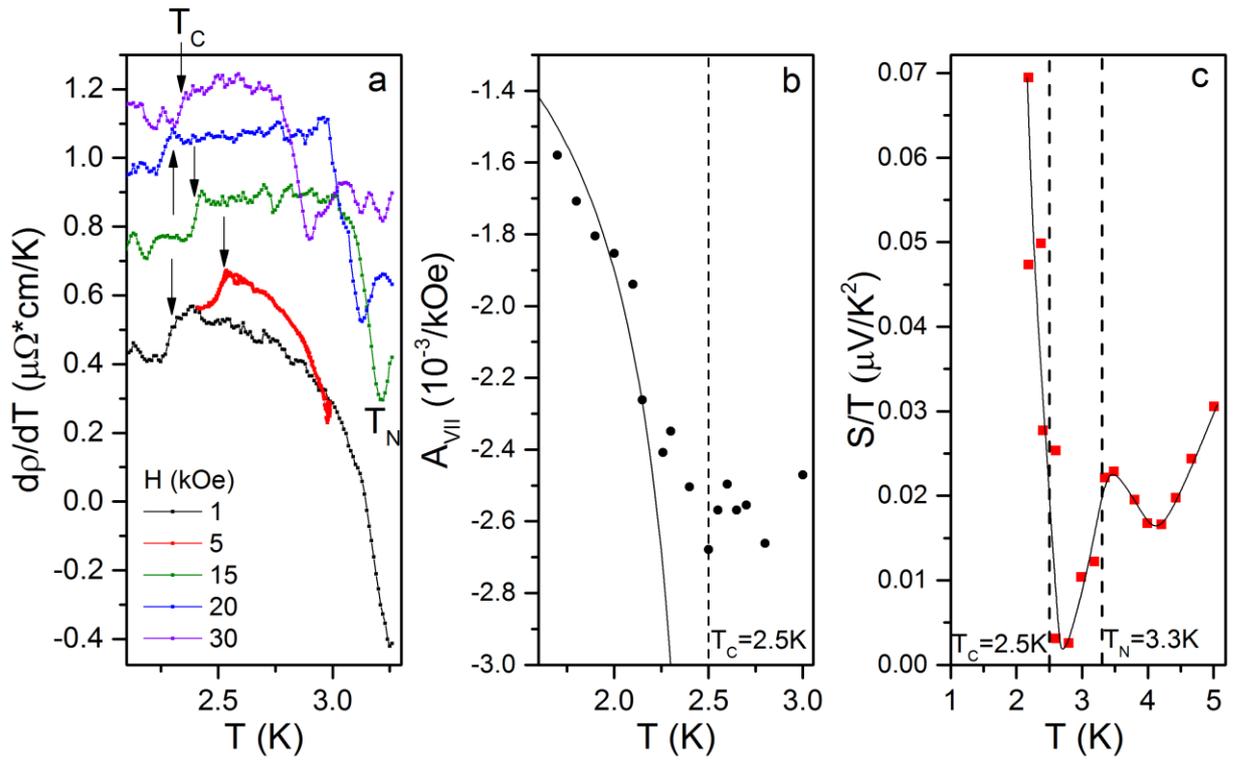

**Fig.S4** (a) Temperature dependence of the linear coefficient in expression Δρ/ρ=A·H for negative magnetoresistance detected in VII phase for **H** ∥ [111]. Line in (b) corresponds to fitting by Eq. (2) with Currie temperature $T_C$ = 2.7 K and an arbitrary scale factor. (c) Temperature dependence of the Seebeck coefficient divided by temperature which demonstrates two features at Neel $T_N$ and Currie $T_C$ temperatures.

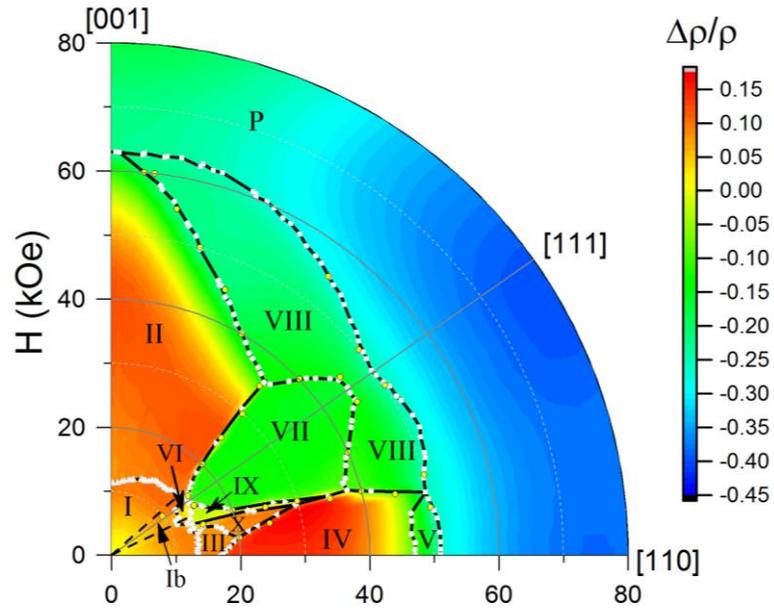

**Fig.S5** Color map of magnetoresistance Δρ/ρ(H,φ) at T = 2 K in the (110) plane with phase boundaries drawn by anomalies on Δρ/ρ(H) and Δρ/ρ(φ) curves. Roman numerals denote the magnetically ordered phases, P denotes the paramagnetic state.